\documentstyle[prl,aps,multicol,epsfig]{revtex}

\begin{document}
\draft
\widetext

%
%
 \title{Electronic Theory for Bilayer-Effects in High-T$_c$ Superconductors}

\author{ S. Grabowski, J. Schmalian, M. Langer and K. H. Bennemann}
\address{ Institut f\"ur Theoretische Physik,
 Freie Universit\"at Berlin, Arnimallee 14, D-14195 Berlin , Germany}

\date{\today}
\maketitle

\widetext
\begin{abstract}
\leftskip 54.8pt
\rightskip 54.8pt
The normal and the superconducting state of two coupled $CuO_{2}$ layers in the High-T$_c$ 
superconductors are investigated by using the bilayer Hubbard model, the FLEX approximation
on the real frequency axis and the Eliashberg theory. We find that the planes are 
antiferromagnetically correlated which leads to a strongly enhanced shadow band formation.
Furthermore, the inter-layer hopping is renormalized which causes a blocking of the quasi 
particle inter-plane transfer for low doping concentrations. Finally, the superconducting
order parameter is found to have a $d_{x^2-y^2}$ symmetry with significant additional 
inter-layer contributions.
 
\end{abstract}


\begin{multicols}{2}   

\narrowtext
 
The importance of the multiple $CuO_{2}$ layers within the High-T$_c$ superconductors like 
Bi$_{2}$Sr$_{2}$CaCu$_{2}$O$_{8+\delta}$ (BSCCO) or YBa$_{2}$Cu$_{3}$O$_{6+\delta}$ (YBCO) 
is intensively studied. Neutron Scattering experiments show clear evidence that 
antiferromagnetically correlated spin fluctuations are quite strong between nearest-neighbor 
layers in YBCO~\cite{tranquada,fong}. Recent angular resolved photoemission (ARPES) experiments 
found indications for two separated bands in YBCO~\cite{campuzano,liu,gofron}, that might be related 
to the existence of two $CuO_{2}$ bands caused by a coherent inter-plane quasi particle transfer. 
However, the small experimental observed bilayer splitting in YBCO and the difficulty to resolve two 
$CuO_{2}$ bands in BSCCO~\cite{ding,shen} support the idea that the strong short ranged 
antiferromagnetic order in the cuprates alters the electronic excitations and reduces the inter-layer 
hopping. Here the observation of shadows of the Fermi surface (FS) in BSCCO~\cite{aebi,chubukov} 
demonstrates impressively that precursors of the magnetic phase transition are already present in the 
normal state. Furthermore, it has been suggested that a reduced quasi particle hopping and a 
Josephson-like coherent Cooper pair tunneling might be responsible for the interesting layer-dependence 
of T$_c$~\cite{anderson}. 

Theoretically, two coupled $CuO_{2}$ layers have been studied microscopically by incorporating solely a 
magnetic coupling~\cite{milis} and within the bilayer Hubbard model including a finite inter-plane 
hopping~\cite{bulut,lichtenstein2}. Here and in recent phenomenological theories~\cite{lichtenstein,levin}, 
it has been argued that the presence of bilayer correlations might lead to a superconducting $s$-wave order 
parameter that has opposite signs in the two $CuO_{2}$ bands ($s^{\pm}$-state). Moreover, for a single layer 
compound with a model dispersion we recently presented results indicating that the shadow states~\cite{aebi} 
are caused by an antiferromagnetic coupling of momenta ${\bf k}$ at the FS and ${\bf k+Q}$ 
(${\bf Q}= (\pi,\pi)$)~\cite{shadow} and studied their importance for the fine structure in tunneling and 
ARPES experiments below $T_c$~\cite{supra}. However, it is {\it a priori} not clear how the realistic 
dispersions of the cuprates and the significant bilayer correlations affect the electronic excitation 
spectrum in BSCCO and YBCO. 

Despite these efforts it remains an open question to determine the influence of the bilayer coupling on the
quasi particle dispersions, on the inter-layer charge dynamics, on c-axis transport properties and most 
of all on the superconducting phase of the cuprates. Here, it might be of particular importance to investigate 
the relation between electronic theories and the results of the model Hamiltonian of Chakravarty {\it et al.}
~\cite{anderson}. 
 
In this letter we study the bilayer Hubbard model within the FLEX approximation~\cite{bickers} on 
the real frequency axis. We find that the planes are antiferromagnetically correlated yielding 
strong deformations of the quasi particle dispersions and a blocking of the effective inter-layer hopping 
for small doping concentrations and low excitation energies. For YBCO- and BSCCO-like systems we observe 
that shadow states occur only when the inter-plane coupling is considered. For the superconducting state 
we find that the $d_{x^2-y^2}$ symmetry is most stable and that a bilayer is characterized by inter- and 
intra-layer Cooper pair formation.
 
Our theory is based on the Hubbard Hamiltonian for a bilayer $CuO_{2}$-system  
\[ 
H=\sum_{i,j,l,l',\sigma} (t^{i,l}_{j,l'}-\mu \; \delta^{i,l}_{j,l'}) \; 
c^{\dagger}_{i,l,\sigma} c_{j,l',\sigma}
+U \sum_{i,l} n_{i,l,\uparrow} n_{i,l,\downarrow}\;,
\]
where the hopping integrals $t^{i,l}_{j,l'}$ determine the bare dispersion 
$\varepsilon_{ll'}^o({\bf k})$ in 2D ${\bf k}$-space, $i$ and $j$ ($l$ and $l'$) are the site 
(layer) indices, $\delta^{i,l}_{j,l'}$ the Kronecker symbol, $U$ the local Coulomb repulsion and 
$\mu$ the chemical potential. The interaction-free contribution of the  Hamiltonian can 
be diagonalized yielding an antibonding ($-$) and a bonding band ($+$) with bare dispersion 
$\varepsilon^o_{\pm}({\bf k})$. Assuming that this symmetry holds also in the full interacting 
case, one can define corresponding electronic Greens function $G_{\pm} ({\bf k},i\omega_m)$ with 
fermionic Matsubara frequencies $\omega_m=(2m + 1)\pi T$ and temperature $T$. They are connected 
to the self energy $\Sigma_{\pm} ({\bf k},i\omega_m)$ via the Dyson equation 
$G_{\pm} ({\bf k},i\omega_m)
=1/(i\omega_m-(\varepsilon^o_{\pm}({\bf k})-\mu)-\Sigma_{\pm} ({\bf k},i\omega_m))$. For the 
FLEX self energy one obtains 
\[
\Sigma_{\lambda} ({\bf k},i\omega_m)= \sum_{k',\lambda'}
V_{\lambda,\lambda'}({\bf k-k'},i\omega_{m}-i\omega_{m'})\;G_{\lambda'} ({\bf k'},i\omega_{m'})
\]
with $\sum_{k'}=\frac{T}{N}\sum_{{\bf k'},m'}\;$, band index $\lambda=\pm$ and number of 
momentum points $N$. Note that due to the symmetry of the bilayer it follows for the intra-band
interactions that $V_{++}({\bf k},i\omega_m)=V_{--}({\bf k},i\omega_m)$ and for the inter-band
contributions that $V_{+-}({\bf k},i\omega_m)=V_{-+}({\bf k},i\omega_m)$. 
$V_{\lambda,\lambda'}({\bf k},i\omega_{m})$ can be obtained by calculating the electron-hole
bubble $\chi_{\lambda,\lambda'}({\bf k},i\omega_m)=-\sum_{k'}
G_{\lambda} ({\bf k'+k},i\omega_m+i\omega_{m'})\;
G_{\lambda'} ({\bf k'},i\omega_{m'})$ and by performing the straightforward summation of the FLEX 
diagrams~\cite{bickers} in the layer representation yielding an inter-plane, 
$V_{\bot}({\bf k},i\omega_m)$, and an in-plane, $V_{\parallel}({\bf k},i\omega_m)$, contribution:
\begin{eqnarray}
V_{++}({\bf k},i\omega_m)&=&
1/2\;(V_{\parallel}({\bf k},i\omega_m)\;+ V_{\bot}({\bf k},i\omega_m))
\nonumber\\ 
V_{+-}({\bf k},i\omega_m)&=&
1/2\;(V_{\parallel}({\bf k},i\omega_m)\;- V_{\bot}({\bf k},i\omega_m))\;.
\label{trafo} 
\end{eqnarray}
\vskip -1.7cm
\begin{figure}
\centerline{\epsfig{file=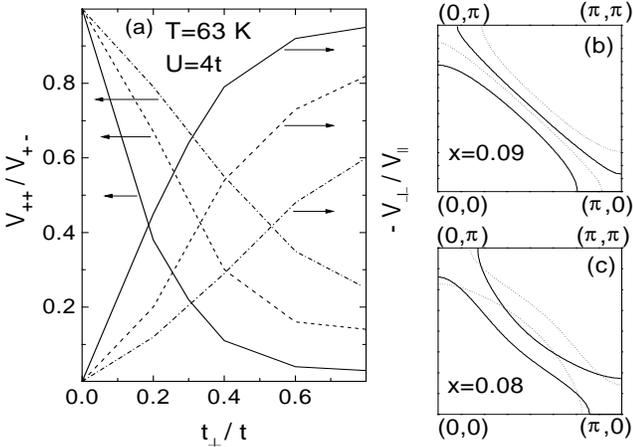,width=9cm,height=11cm}}
\vskip -3.7cm
\caption{Dependence of the effective interactions on the doping and on the inter-layer
hopping $t_{\bot}$. (a) Ratio -V$_{\bot}$/V$_{\parallel}$ and V$_{++}$/V$_{+-}$ for the 
LSCO-like model. $x=0.09$ (solid line), $x=0.12$ (dashed line) and $x=0.16$ (dashed-dotted line).
Fermi surfaces and their shadows of the LSCO- (b) and YBCO-like model (c) dispersions.}
\label{fig1}
\end{figure}
This set of coupled equations is solved self-consistently on the real frequency axis by using
the numerical method presented in Ref.~\cite{details}. To characterize the FS of the different
High-T$_c$ superconductors we use the bare dispersion
\begin{eqnarray}
\varepsilon^o_{\pm}({\bf k})&=&-[2t(\cos (k_x)+\cos (k_y)) + 4t'\cos (k_x) \cos (k_y)
\nonumber\\
&& +2t'' (\cos (2k_x)+\cos (2k_y))\pm t_{\bot}] 
\nonumber
\end{eqnarray}
with hopping integrals $t=0.25$ eV, $t'=-0.38\;t$ and $t''=-0.06\;t$ as model for YBCO
~\cite{radtke,parameter} and compare these results with a model dispersion for $t'=t''=0$. Since the later 
FS is similar to the La$_{2-x}$Sr$_{x}$CuO$_{4}$ system for $t_{\bot}=0$, we call this dispersion for 
simplicity LSCO-like although LSCO has in distinction to YBCO or BSCCO only one plane per unite cell. We 
investigate the complete dependence of the electronic properties on $t_{\bot}$, but take 
$t_{\bot}=0.4\;t = 100$ meV as inter-plane hopping in YBCO~\cite{andersen}. For the comparison with previous 
results we take $U=4t$ but notice that we find no significant changes in our data up to values of $U = 6t$.  
  
In Fig. 1(a) we investigate the interplay between the inter- and intra-plane antiferromagnetic correlations 
in bilayer cuprates upon $t_{\bot}$ and the doping $x$, where $n=1-x$ is the occupation number per site. By 
using the LSCO-like dispersion we find that $V_{\parallel}({\bf k},\omega)$ decreases only slightly when 
$t_{\bot}$ is increased and that the magnetic in-plane correlation length $\xi$~\cite{shadow} is almost not 
sensitive to the inter-layer hopping~\cite{YBCO}. Since $V_{\bot}({\bf k},\omega)$ and 
$V_{\parallel}({\bf k},\omega)$ have the same ${\bf k}$-dependence, but with opposite signs, the two layers 
are antiferromagnetically correlated as observed in experiments~\cite{tranquada,fong}. As a measure of this 
phenomenon we plot in Fig. 1(a) the ratio of the maxima of the corresponding interactions 
$-V_{\bot}/V_{\parallel}:=-V_{\bot}({\bf Q},\omega)/V_{\parallel}({\bf Q},\omega)$.
Here we observe that the bilayer coupling increases rapidly with increasing $t_{\bot}$ until it saturates. 
In addition, it is remarkable that not only  $V_{\bot}({\bf k},\omega)$ and $V_{\parallel}({\bf k},\omega)$ 
increase with decreasing doping, but also the ratio $-V_{\bot}/V_{\parallel}$. Thus, the stabilization of 
the inter-plane magnetism via a coupling of antiferromagnetically ordered in-plane regions (size $\sim \xi$) 
across the layers should be most effective for low doping concentrations. By transforming these results to 
the band representation via Eq.~\ref{trafo}, one sees in Fig. 1(a), where we plot 
$V_{++}/V_{+-}:=V_{++}({\bf Q},\omega)/V_{+-}({\bf Q},\omega)$, that the contribution of the intra-band 
interactions vanishes rapidly for low doping with increasing $t_{\bot}$ thereby explaining the absence of 
these excitations in experiments~\cite{tranquada}. The suppression of $V_{++}({\bf Q},\omega)$ and the 
corresponding enhancement of $V_{+-}({\bf Q},\omega)$ can also be understood by considering the FS topology 
of the LSCO- and YBCO-like models in Fig. 1(b) and (c), where the bonding and the antibonding FS 
are almost shifted by ${\bf Q}$ and the fact that we find very flat quasi particle bands at ($\pi,0$).
Finally, the later point leads also to an antibonding FS that is closed around ($0,0$), which is 
in contrast to LDA calculations~\cite{andersen} and might also be interesting for the current debate about
the FS in YBCO~\cite{shen}.
 
The dominance of the inter-layer and inter-band correlations has important consequences for the quasi particle 
dispersions and the bilayer splitting. By investigating the bonding and the antibonding bands of the LSCO model
with $t_{\bot}=0.4\;t$~\cite{YBCO}, we observe similar to Ref.~\cite{lichtenstein2} that the bilayer splitting is 
strongly renormalized from the uncorrelated $\Delta \varepsilon({\bf k}=(\pi,0))=2\;t_{\bot}=200$ meV to
$\Delta \varepsilon({\bf k}=(\pi,0))=125$ meV in the case of $x=0.12$. To study the origin of this effect 
and more interestingly to investigate the low energy inter-plane electron dynamics 
${\rm Re}\;\Sigma_{\bot}({\bf k},\omega)=0.5\;{\rm Re}\;(\Sigma_{+}({\bf k},\omega)-\Sigma_{-}({\bf k},\omega))$
is plotted in Fig. 2 for ${\bf k}=(\pi,0)$ upon doping. This quantity renormalizes $t_{\bot}$ and yields an 
effective frequency and momentum dependent inter-plane hopping amplitude 
$\tilde{t}_{\bot}({\bf k},\omega)=t_{\bot}-{\rm Re}\;\Sigma_{\bot}({\bf k},\omega)$, which follows directly from 
the matrix Dyson equation in the layer representation. The sign of ${\rm Re}\;\Sigma_{\bot}({\bf k},\omega)$ is 
determined for small $\omega$ by the sign of $V_{\bot}({\bf k},\omega)$, such that we find 
$\tilde{t}_{\bot}({\bf k},\omega)\ll t_{\bot}$ for $\omega<t_{\bot}$ and low $x$ ($x<0.12$). This interesting 
phenomenon can be found along the entire FS although it is most pronounced at ($\pi,0$) and exists for a wide range 
of $t_{\bot}$ as can be observed in the inset of Fig. 2. Consequently, the effective inter-plane hopping at the 
Fermi energy is blocked in the cuprates, which leads to the disappearance of a coherent quasi particle tunneling 
process although due to the frequency dependence of ${\rm Re}\;\Sigma_{\bot}({\bf k},\omega)$ both bands are still 
splitted. Physically, this is related to the fact that each inter-layer hopping process is accompanied by spin-flips
due to the opposite antiferromagnetic environment in the other plane. Thus, due to the corresponding large 
${\rm Im}\;\Sigma_{\bot}({\bf k},\omega)$ an incoherent coupling of the layers occur. Note that a hypothetical 
ferromagnetic coupling across the bilayers would not suppress $\tilde{t}_{\bot}({\bf k},\omega)$, since there is no 
competition between the kinetic energy gain and the magnetic correlations. Finally, the remarkable doping dependence 
of our results might be of particular importance for the $c$-axis transport properties of the cuprates, since 
{\it underdoped} and {\it overdoped} compounds are characterized by qualitatively different temperature 
dependences~\cite{levin2}.
\vskip -0.7cm
\begin{figure}
\centerline{\epsfig{file=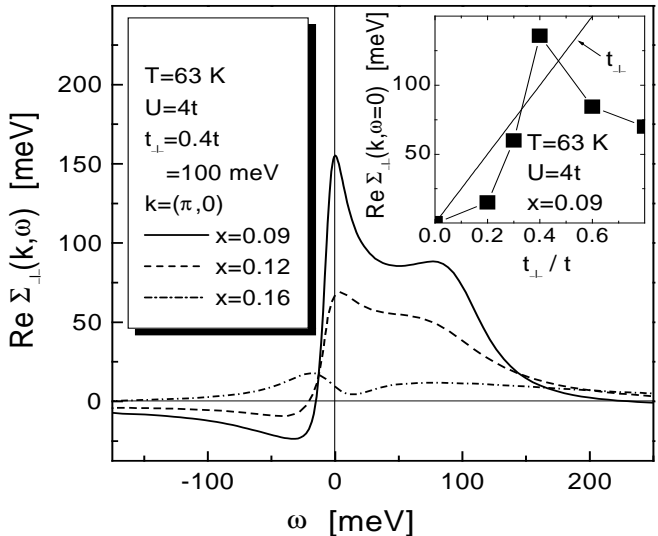,width=9cm,height=11cm}}
\vskip -3.3cm
\caption{${\rm Re}\;\Sigma_{\bot}({\bf k},\omega)$ for ${\bf k}=(\pi,0)$ and for different doping values. Inset:
${\rm Re}\;\Sigma_{\bot}({\bf k},0)$ at the FS of the antibonding band versus $t_{\bot}$, where $\omega=0$ refers 
to the Fermi energy.}
\label{fig2}
\end{figure}
In Fig. 3 we demonstrate the influence of the inter-plane coupling on the formation of shadow states. In 
Ref.~\cite{shadow} we showed for a single layer system that the antiferromagnetic coupling of momenta 
${\bf k}$ and ${\bf k+Q}$ leads to a transfer of spectral weight from the FS to its shadow. Now, since we find 
for a bilayer system that $V_{++}({\bf k},\omega)$ is small compared to $V_{+-}({\bf k},\omega)$ for intermediate 
$t_{\bot}$, we expect that the spectral weight is not only shifted by the momentum ${\bf Q}$, but simultaneously 
also from the bonding to the antibonding band and vise versa. This phenomenon is demonstrated in Fig. 3 for the 
LSCO-like model dispersion, where we present our results for the spectral density for the bonding 
($\varrho_{+} ({\bf k}_{FS},\omega)$) and the antibonding ($\varrho_{-} ({\bf k}_{FS}+{\bf Q},\omega)$) band and FS 
momentum ${\bf k}_{FS}$. There is a transfer from the bonding to the antibonding band leading to an occupied shadow 
state and in addition a transfer in the reversed direction above the Fermi energy. Most interestingly, the intensity 
of the shadow states is strongly dependent on the magnitude of the inter-plane hopping as can be observed in the 
right inset of Fig. 3. 
\vskip -0.7cm
\begin{figure}
\centerline{\epsfig{file=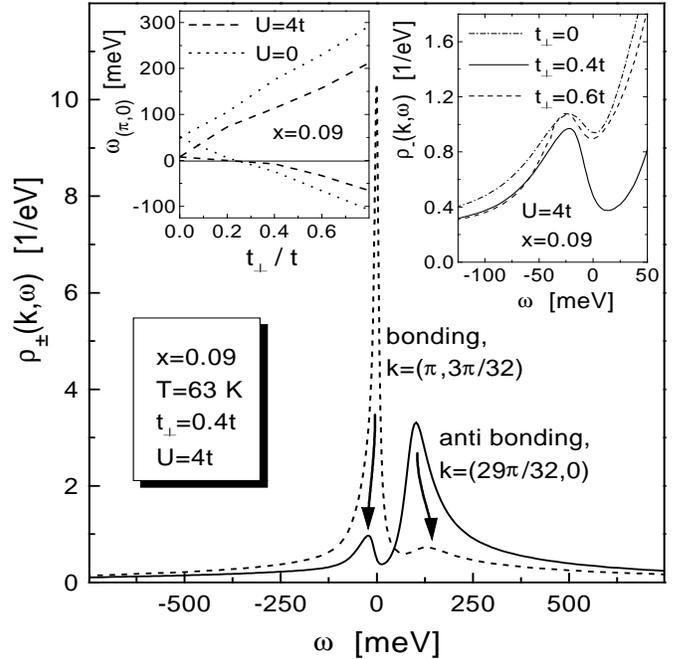,width=9cm,height=11cm}}
\vskip -1.4cm
\caption{Spectral density $\varrho_{\pm} ({\bf k}_{FS},\omega)$ for the LSCO-like model. 
Right inset: Dependence of the shadow state intensity on $t_{\bot}$.
Left inset: Position of the bands at the ($\pi,0$) point with respect to the Fermi energy 
for $U=0$ and $U\neq0$.}
\label{fig3}
\end{figure}
Here the shadow peak is most pronounced for $t_{\bot}=0.4\;t$ before it decreases again for even larger values 
of $t_{\bot}$. To verify that this result is related to an enhanced antiferromagnetic coupling, we 
investigated ${\rm Im}\;\Sigma_{-}({\bf k}_{FS}+{\bf Q},0)$ as a measure of the coupling of the FS to its 
shadow~\cite{shadow} and found a corresponding maximum at $t_{\bot}=0.4\;t$. Therefore, the strongly enhanced
formation of shadow states in bilayers is caused by electronic correlations and is related to the
already discussed suppression of the bilayer splitting and a corresponding large spectral density at the 
Fermi level. This is demonstrated in the left inset of Fig. 3, where we plot the $t_{\bot}$-dependence of 
the position ($\omega_{(\pi,0)}$) of the flat quasi particle band at ($\pi,0$) with respect to the Fermi level 
for $U\neq0$ and $U=0$. It can be seen that the antibonding band moves away from the Fermi energy 
($\sim t_{\bot}$), while the bonding band is tightly pinned at $\omega=0$ up to $t_{\bot}\approx 0.45 \;t$. 
Note that the shadow state intensity is also enhanced for larger $x$ and most importantly that in a bilayer 
shadow states can be found up to doping concentrations of $x \approx 0.17$, while they vanish for 
$x \approx 0.13$ in a single layer compound.
  
By performing calculations for the YBCO dispersion we find no shadow states for $t_{\bot}=0$. Therefore, 
due to the similarity of the YBCO and BSCCO FS~\cite{bscco} the experimental observation by Aebi {\it et al.}
~\cite{aebi} can not be satisfactorily understood by considering only a single $CuO_{2}$ plane. However, by taking 
a finite $t_{\bot}$ into account we find for YBCO in agreement with our results for the LSCO-like model that 
the antiferromagnetic coupling increases and that shadow states start to appear for $x<0.12$ with a maximum 
intensity at $t_{\bot} \approx 0.4\;t$. Furthermore we find that the most favorable region to observe shadow 
states in ARPES is in the neighborhood of the ($\pi/2,\pi/2$) point, where main and shadow band are well 
separated. Near ($\pi,0$) the absolute intensity of the shadow states is largest, but they are difficult to 
detect because of the superposition of shadow peaks and the dominant main band contributions~\cite{bscco}.
\vskip -1.0cm
\begin{figure}
\centerline{\epsfig{file=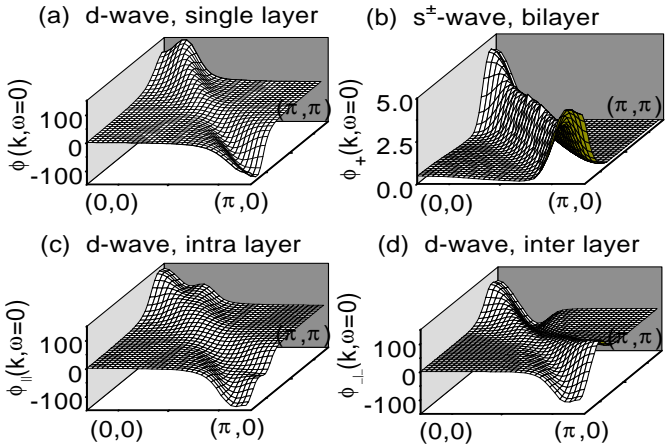,width=9cm,height=11cm}}
\vskip -4.3cm
\caption{Superconducting order parameter for $\omega=0$, $T=48$ K and $U=4t$ (in meV). 
(a) single layer ($t_{\bot}=0$). 
(b) meta-stable bilayer $s^{\pm}$-state for the bonding band and $t_{\bot}=0.4\;t$.
(c-d) bilayer $d$-wave state in the layer representation for  $t_{\bot}=0.4\;t$.}
\label{fig4}
\end{figure}
The superconducting state of the bilayer Hubbard model was treated by using a strong coupling Eliashberg 
theory within the fully self-consistent FLEX approximation~\cite{detailssup}. By assuming that there is no 
inter-band pairing, one obtains one order parameter for each band, namely $\phi_{\pm}({\bf k},\omega)$, 
which are connected to the layer representation via 
$\phi_{\pm}({\bf k},\omega)=\phi_{\parallel}({\bf k},\omega)\pm\phi_{\bot}({\bf k},\omega)$.
By solving these equations for the YBCO dispersion with $t_{\bot}=0$, we find for all $x$ a superconducting 
state with $d_{x^2-y^2}$ pairing symmetry and values of $T_c$ ($T_c=64$ K for $x=0.12$) that are only slightly 
smaller than the $T_c$ of the previously investigated LSCO-like dispersion ($T_c=78$ K)~\cite{supra}. 
Concerning the pairing symmetry in a bilayer system, we expect in view of our results for the effective 
interactions for larger doping an intra-plane $d_{x^2-y^2}$ state. However, since the inter-band interactions 
are dominant for low doping concentrations, the argumentation of Ref.~\cite{lichtenstein} might apply leading 
to an $s^{\pm}$-symmetry, which corresponds to an inter-plane pairing state. To clarify this point, we present
in Fig. 4 our results for $T$ well below $T_c$ and the LSCO-like model with $t_{\bot}=0.4\;t$ and $x=0.09$ in 
comparison with $t_{\bot}=0$~\cite{YBCO}. Here in contrast to the single layer case, where only a $d$-wave 
pairing state occurs (Fig. 4(a)), we find an $s^{\pm}$-state as a solution of the Eliashberg equations as shown 
in Fig. 4(b). Note, this order parameter is highly anisotropic and largest at the FS to achieve the highest 
energy gain. Nevertheless, the $s^{\pm}$-state refers only to a meta-stable solution, since the bilayer 
$d_{x^2-y^2}$ pairing symmetry, as shown in Fig. 4(c) and (d), still yields $\phi^{d} \gg \phi^{s}$ and 
consequently a much larger condensation energy. Interestingly, the $d$-wave state is characterized by an 
increasing contribution of inter-layer pairing with decreasing doping, where for $x=0.09$ Cooper pairs are 
formed by electrons from the same layer as from different layers with almost equal probability. However, we 
did not find a large enhancement of the corresponding $T_c$, which might be caused by magnetic frustrations,
since the inter-layer $d$-wave pairing occurs also for lattice sites that are ferromagnetically coupled above 
$T_c$.
 
In conclusion, we presented new results for the doping dependence of the bilayer Hubbard model by using the 
FLEX approximation on the real frequency axis. For realistic inter-plane hopping integrals we find that the 
antiferromagnetic coupling across the layers is strong in YBCO, which leads to an enhancement of the shadow 
state intensity and to the appearance of shadow states in YBCO- and BSCCO-like systems. Furthermore, the
short ranged antiferromagnetic order reduces the bilayer splitting and blocks the inter-layer hopping for 
low doping and small excitation energies. In the superconducting state we found an order parameter with
$d_{x^2-y^2}$ symmetry that has equally important inter- and intra-layer pairing contributions. However, it 
remains an interesting problem to analyze the observed increase of $T_c$ with the number of layers within one 
unit cell. This might be achieved by including intrinsic Josephson tunneling into our description of the inter-layer
$d$-wave state~\cite{anderson}.
 
We acknowledge the financial support of the DFG and thank Z.X. Shen for sending us his papers prior
to publication.
 
%
%
%
%

  
\end{multicols}

\begin{references}
\bibitem{parameter} 
This model dispersion is used, since its FS is closed around ($\pi,\pi$) for $U=t_{\bot}=0$  
and because of the observed flat quasi particle bands. The later point was achieved by taking 
$t''=-0.06\;t$ instead of $t''=0.06\;t$ as suggested by Ref.~\onlinecite{radtke}. However, our 
results do not depend sensitively on the particular choice of the hopping integrals.
\bibitem{YBCO}
We can draw similar conclusions for the effective interactions, for the bilayer splitting, for the 
shadow states and for the superconducting state from the YBCO dispersion.
\end{references}
\end{document}